# Autonomous Voltage Control for Grid Operation Using Deep Reinforcement Learning

Ruisheng Diao, Zhiwei Wang, Di Shi, Qianyun Chang, Jiajun Duan, Xiaohu Zhang
GEIRI North America
San Jose, CA, USA
Ruisheng.Diao@geirina.net

*Abstract*—Modern power grids are experiencing grand challenges caused by the stochastic and dynamic nature of growing renewable energy and demand response. Traditional theoretical assumptions and operational rules may be violated, which are difficult to be adapted by existing control systems due to the lack of computational power and accurate grid models for use in real time, leading to growing concerns in the secure and economic operation of the power grid. Existing operational control actions are typically determined offline, which are less optimized. This paper presents a novel paradigm, Grid Mind, for autonomous grid operational controls using deep reinforcement learning. The proposed AI agent for voltage control can learn its control policy through interactions with massive offline simulations, and adapts its behavior to new changes including not only load/generation variations but also topological changes. A properly trained agent is tested on the IEEE 14-bus system with tens of thousands of scenarios, and promising performance is demonstrated in applying autonomous voltage controls for secure grid operation.

## I. INTRODUCTION

With the fast-growing penetration of renewable energies, distributed energy resources, demand response and new electricity market behavior, conventional power grid with decades-old infrastructure is facing grand challenges such as fast and deep ramps and increasing uncertainties (e.g., the Californian duck curves), threatening the secure and economic operation of power systems. In addition, traditional power grids are designed and operated to withstand N-1 (and some N-2) contingencies, required by NERC standards [1]. Under extreme conditions, local disturbances, if not controlled properly, may spread to neighborhood areas and cause cascading failures, eventually leading to wide-area blackouts. It is therefore of critical importance to promptly detect abnormal operating conditions/stress events, understand the growing risks and more importantly, apply timely and effective control actions to bring the system back to normal after large disturbances.

Automatic controllers including excitation system, governors, PSS, AGC, etc., are designed and equipped for generator units to maintain voltage and frequency profiles once a disturbance is detected [2]. However, manual actions from system operators are still needed on a daily routing to mitigate operational risks that cannot be handled by the existing automatic controls because of the complexity and high dimensionality of modern power grid. These actions include generator re-dispatch deviating from their scheduled operating points, switching capacitors and other shunt elements, shedding loads under emergency conditions, reducing critical path flows, tripping generators, and so on. The time of application, duration and size of these manual actions are typically determined offline by running massive simulations considering the projected "worst" operating scenarios and contingencies, in forms of decision tables and operational orders. Due to the dynamic and stochastic nature of power systems, it is very difficult to precisely estimate future operating conditions and to predict the control performances, leading to the fact that the offline determined control strategies are either too conservative (causing over investment) or risky (causing stability concerns) when applied in real world.

To fill such reality gaps, several measures have been deployed by power utilities and independent system operators (ISOs). Performing security assessment in near real time is one example, which can effectively understand the operational risks if a contingency occurs [3]. However, the lack of computing power and sufficiently accurate grid models prevents optimal control actions from being derived and deployed in real time. Machine learning based methods, e.g., decision trees, support vector machines, neural networks, have also been proposed to first train agents using offline analysis and then apply in real time [4]-[6]. Majority of these approaches focus only on monitoring and security assessment, rather than performing and evaluating controls for operation. Consequently, this paper presents a novel autonomous control paradigm, called **Grid Mind**, for grid operation that takes advantage of the successful artificial intelligent (AI) technology, deep reinforcement learning (DRL), to derive fast and effective controls in real time targeting at the current and near-future operating conditions. Using DRL for intelligent controls has achieved significant success in various domains, such as AlphaGo [7], ATARI Games [8], robotics [9], and demand response to improve energy efficiency [10]. Applying the most recent DRL research outcomes for autonomous controls in power grid daily operation has been rarely reported. This work aims at bridging this gap by developing a measurement-driven, autonomous and self-evolving voltage controller for grid operation using DRL.

The remainder of this paper is organized as follows. Section II provides the architecture design of the proposed **Grid Mind** framework, the principles of DRL and the autonomous voltage control flowchart. Section III discusses the in-depth implementation. In Section IV, case studies are conducted using the developed tool, which shows excellent performance on the IEEE 14-bus test system. Finally, conclusions are drawn in Section V and future research work is also identified.

## II. THE PROPOSED AUTONOMOUS VOLTAGE CONTROL SCHEMA FOR GRID OPERATION USING DRL

### A. Architecture Design

The architecture design of **Grid Mind** is depicted in Fig.1, where the **Grid Mind** AI agent is trained offline by interacting with massive offline simulations and historical events. Once abnormal conditions are sensed in real time, the agent will provide suggested actions and the corresponding expected results. The control actions will be firstly verified by human operators before actually implemented, to enhance robustness and guarantee performance. After the action has been taken by the power grid (environment) at the current state, the agent will receive a reward from the environment containing the next set of states, to evaluate the effectiveness of policy. In the meantime, the relationship among action, states and reward are updated in the Agent's memory. This process continues as the Agent keeps learning and improving its performance over time.

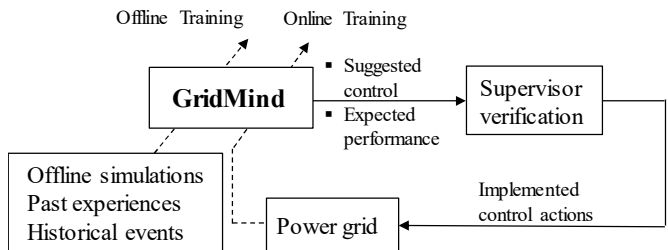

Fig.1. Concept of the **Grid Mind** framework -- autonomous controls for grid operation using DRL.

### B. Principles of DRL

Artificial Intelligence is a process when computers try to solve specific tasks or problems by mimicking human's behavior; and machine learning (ML) is a subset of AI technologies by learning from data or observations and then making decisions based on trained models [11]. ML consists of supervised learning, unsupervised learning, semi-supervised learning and reinforcement learning (RL), serving different purposes. Different from all other branches, RL refers to an agent that learns its action policy that maximizes the expected rewards based on interactions with environments. Typical RL algorithms include dynamic programming, Monte Carlo and Temporal difference such as Q-learning [12]. An RL agent continuously interacts with an environment; where the **environment** receives an action, emits new states and calculates a reward; and the **agent** observes states, suggests action to maximize next reward. Training an RL agent involves dynamically updating a policy (mapping from states to action), a value function (mapping from action to reward) and a model (for representing the environment).

Deep learning (DL) provides a general framework for representation learning that consists of many layers of nonlinear functions mapping inputs to outputs. Its uniqueness rests with the fact that DL does not need to specify features beforehand. One typical example is the deep neural network [12]. Basically, DRL is a combination of DL and RL, where DL is used for representation learning and RL for decision making. In the proposed **Grid Mind** framework, deep Q network (DQN) [15] is used to estimate the value function, which supports continuous state sets and is suitable for power grid control. The designed DRL agent in the framework for autonomous voltage control is shown in Fig.2. More details regarding each component are provided in the subsequent sections.

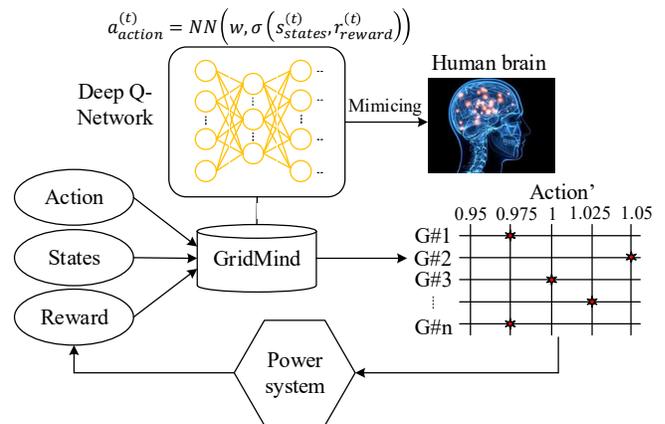

Fig. 2. Designed RL agent for autonomous grid control.

### C. Autonomous Voltage Control using DRL Agent with Practical Considerations

The goal of a well-trained DRL agent for autonomous voltage control is to provide an effective action from finite control action sets when observing abnormal voltage profiles. The definition of episode, states, action and reward is given below:

#### 1) Episode

An episode represents any operating condition collected from real-time measurement systems such as supervisory control and data acquisition (SCADA) or phasor measurement unit (PMU), under random load variations, generation dispatches, topology changes and contingencies. In this work, only quasi-steady state is considered without considering transient behaviors in between. Contingencies are randomly selected and applied in this work to mimic reality.

#### 2) States

The states are defined as a vector of system information that is used to represent system conditions, including active and reactive power flows on transmission lines and transformers, as well as bus voltage magnitudes and phase angles.

#### 3) Action Space

Typical manual control actions to mitigate voltage issues include adjusting generator terminal voltage setpoints, switching shunt elements, transformer tap ratios, etc. In this work, without loss of generality, we consider generator voltage set point adjustments as actions to maintain system voltage profile. Each generator voltage setpoint can be adjusted within a range, e.g., [0.95, 0.975, 1.0, 1.025, 1.05] p.u. The combination or permutation of all available generator setpoints forms an action space used to train a DRL agent.

*4) Reward*

Several voltage operation zones are defined to differentiate voltage profiles, including normal zone (0.95-1.05 pu), violation zone (0.8-0.95 pu or 1.05-1.25 pu) and diverged zone (>1.25 pu or <0.8 pu), as shown in Fig.3.

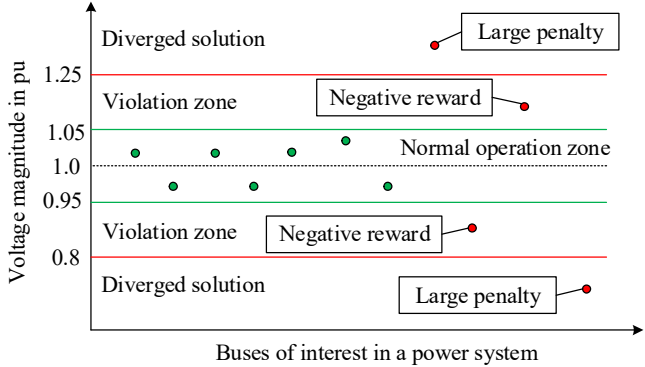

Fig. 3. Voltage profile zone definition for calculating rewards.

Rewards are designed accordingly for each zone. In one episode (Ep), define $V_i$ as the voltage magnitude at bus $i$, and the reward for the $j^{th}$ control iteration can be calculated as:

$$Reward_j \begin{cases} \text{large reward}(+100), \forall V_i \in \text{normal opertaion zone} \\ \text{large penalty}(-100), \exists V_i \in \text{diverged zone} \\ \text{negtive reward}(-50), \exists V_i \in \text{violoation zone} \end{cases} \quad (1)$$

The final reward for the entire episode containing $n$ iterations is then the total accumulated rewards divided by the number of control iterations as

$$\text{Final Reward} = \sum_{j=1}^{n} Reward_j / n \quad (2)$$

In this way, a higher reward is assigned to very effective action (taking one action only vs many action iterations) to solve the same voltage problem. With the above definition of DRL components, the computational flowchart of training a DRL agent is given in Fig. 4, which consists of several key steps:

**Step 1**: starting from one episode (real-time information collected in a power network), solve power flow and check potential voltage violations. A typical violation range can be defined as 0.95-1.05 p.u. for all buses of interest in the power system being studied;

**Step 2**: based on the states obtained, a reward value can be calculated, both of which are fed into the DRL agent; the agent then generates an action based on its observation of the current states and expected future rewards;

**Step 3**: the environment (e.g., AC power flow solver) takes the suggested action and solve another power flow. Then, bus voltage violations are checked again. If no more violation occurs, calculate the final reward for this episode and terminate the process of the current episode;

**Step 4**: if the violation is detected, check for divergence. If so, update the final reward and terminate an episode. If power flow converges, evaluate reward and return to **Step 2**.

The training process terminates when one of the three conditions is met: (1) no more violation occurs, (2) power flow diverges, or (3) the maximum number of iterations is reached.

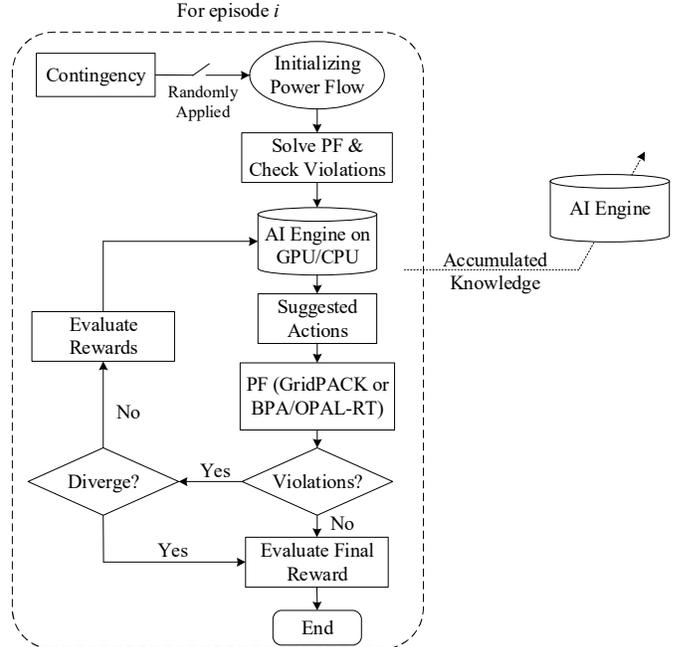

Fig. 4. Computational flowchart of training a DRL Agent for autonomous voltage control under contingencies.

## III. Implementation of DRL Agent

### A. Platform

The platform used to train and test DRL agents for autonomous voltage control is CentOS 7 Linux Operation System (64 bit). This server is equipped with Intel Xeon E7-8893 v3 CPU at 3.2 GHz and 528 GB memory. All the DRL training and testing process are performed on this platform.

### B. Power Grid Simulator

To mimic real power system environment, a power grid simulator developed by the Pacific Northwest National Laboratory, GridPACK$^{TM}$, is adopted, which is equipped with function modules such as power flow, dynamic simulation, contingency analysis, state estimation and so on [13]. In this work, only the AC power flow module is applied to interact with the DRL Agent. Intermediate files are used to pass information between GridPACK and the DRL Agent, including power flow information file saved in PTI v33 raw format and power flow solution results saved in text files.

### C. DRL Agent

For DRL Agent, the most recently developed DQN libraries in Anaconda is utilized, which is a popular python data science platform for implementing AI technologies [14]. This platform provides useful libraries including Keras, Tensorflow, Numpy and others for effective DQN agent development. The Deep Q-learning framework introduced in [15] is also used to set up the environment of DRL Agent and to interact with GridPACK, which is coded using Python 3.6.5 scripts. The information flow is given in Fig. 5.

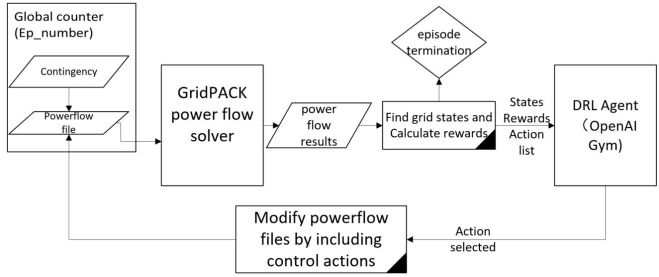

Fig. 5. Information flow of the DRL Agent training process.

## IV. CASE STUDIES AND DISCUSSION

The proposed DRL Agent for autonomous voltage control is tested on the IEEE 14 bus system model, with 14 buses, 5 generators, 11 loads, 17 lines and 3 transformers. The total system load is 259 MW and 73.5 MVAr. A single-line diagram of the system is shown in Fig. 6. To test the performance of the DRL agent, massive operating conditions to mimic reality are created and three case studies are conducted.

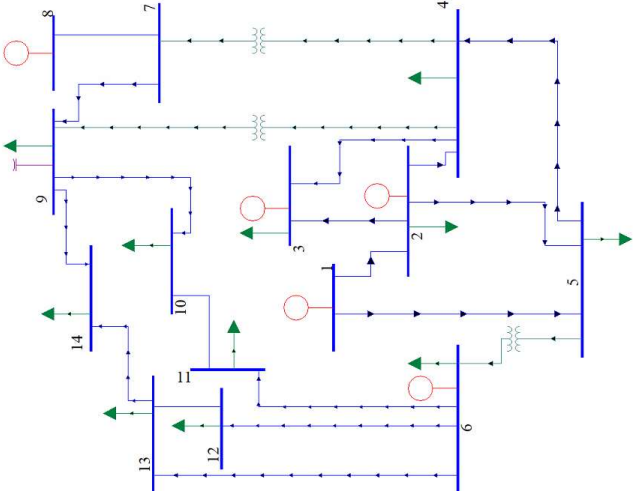

Fig.6. Single-line diagram of the IEEE 14 bus system.

### A. Case I – Without Contingencies

In Case I, all lines and transformers are in service without any topology changes. Random load changes are applied across the entire system, and each load fluctuates within 80%-120% of its original value. When loads change, generators are re-dispatched based on a participation factor list to maintain system power balance. The commercial software package, Powerflow & Short circuit Assessment Tool (PSAT) developed by Powertech Labs in Canada, is used to generate 10,000 random operating conditions based on above rules. Each case is a converged power flow with or without voltage violations (normal range: 0.95-1.05 pu), saved in PTI v33 format files. A DRL Agent is trained using the proposed method and its performance on the 10,000 episodes is shown in Fig. 7. The x-axis represents the number of episodes being trained; while y-axis represents the calculated final reward values. It can be observed that the rewards of the first few hundreds of episodes are relatively low, given that the Agent starts with no knowledge about controlling the voltage profiles of the grid. As the learning process continues, the Agent takes fewer and fewer control actions to fix voltage problems. It is worth mentioning that several parameters in the DQN agent play a role in deciding when to explore new random actions versus using existing models. These parameters include exploration rate, learning speed, decay and others, which need to be carefully tuned to achieve satisfactory performance. In general, when the Agent performs well on a large number of unseen episodes, one can trust the trained model more and use it for online applications.

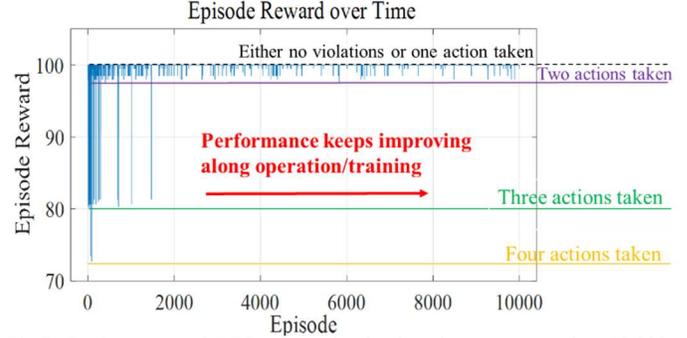

Fig.7. Performance of DRL Agent in the learning process using 10,000 episodes without considering contingencies.

Table I explains the details of the Agent's intelligence in Episode 8 and 5000. For the initial system condition in Episode 8, several bus voltage violations are identified, shown in the first row of Table I. To fix the voltage issues, the Agent took an action by setting generator voltage setpoint to [1.05 1.025 1 0.95 0.975] for the 5 generators; after this action, the system observes less violations, shown in the second row of Table I. Then, the Agent took a second action [1.025 0.975 0.95 1 1.05] before all the voltage issues are fixed. By the time the Agent learns 4999 episodes, it accumulates sufficient knowledge: at the initial condition of Episode 5000, 6 bus voltage violations are observed, highlighted in the 4th row of Table I. The Agent took one action and corrected all voltage issues, using the policy that DQN memorizes.

TABLE I
SYSTEM STATES OF EP 8 AND EP 5000, IN CASE I

| bus1 | bus2 | bus3 | bus4 | bus5 | bus6 | bus7 | bus8 | bus9 | bus10 | bus11 | bus12 | bus13 | bus14 | episode |
|---|---|---|---|---|---|---|---|---|---|---|---|---|---|---|
| 1.06 | 1.045 | 1.01 | 1.01797 | 1.02025 | 1.07 | 1.06204 | 1.09 | 1.05682 | 1.05137 | 1.05756 | 1.05568 | 1.05237 | 1.03698 | 8 |
| 1.05 | 1.025 | 1 | 0.97375 | 0.9756 | 0.95 | 0.974 | 0.975 | 0.96352 | 0.95255 | 0.94802 | 0.93591 | 0.9342 | 0.93076 | 8 |
| 1.025 | 0.975 | 0.95 | 0.95572 | 0.95909 | 1 | 1.00554 | 1.05 | 0.99402 | 0.98678 | 0.99011 | 0.98523 | 0.98225 | 0.96972 | 8 |
| 1.06 | 1.045 | 1.01 | 1.01936 | 1.01936 | 1.07 | 1.06047 | 1.09 | 1.05409 | 1.04913 | 1.05583 | 1.05456 | 1.05036 | 1.03339 | 5000 |
| 0.975 | 1 | 0.95 | 0.9627 | 0.96341 | 1.025 | 1.01158 | 1.05 | 1.00331 | 0.99898 | 1.00803 | 1.00845 | 1.00369 | 0.9834 | 5000 |

DRL AGENT ACTIONS FOR EP 8 AND EP 5000, IN CASE I

| gen1_vset | gen2_vset | gen3_vset | gen6_vset | gen8_vset | episode |
|---|---|---|---|---|---|
| 1.05 | 1.025 | 1 | 0.95 | 0.975 | 8 |
| 1.025 | 0.975 | 0.95 | 1 | 1.05 | 8 |
| 0.975 | 1 | 0.95 | 1.025 | 1.05 | 5000 |

### B. Case II – With Contingencies

In Case II, the same number of episodes are used, but random N-1 contingencies are considered to represent emergency conditions in real grid operation. Several line outages are considered in this study, 1-5, 2-3, 4-5, 7-9. Each episode picks one line outage randomly, before feeding into the learning process. Shown in Fig. 8, the DRL Agent performs very well when testing on these episodes with random contingencies. Initially, the Agent never meets the episodes with contingencies before and thus takes more actions to fix voltage profiles. After several hundreds of trials, it can fix the voltage profiles using less than two actions for most of the episodes, which demonstrate its excellent learning capabilities.

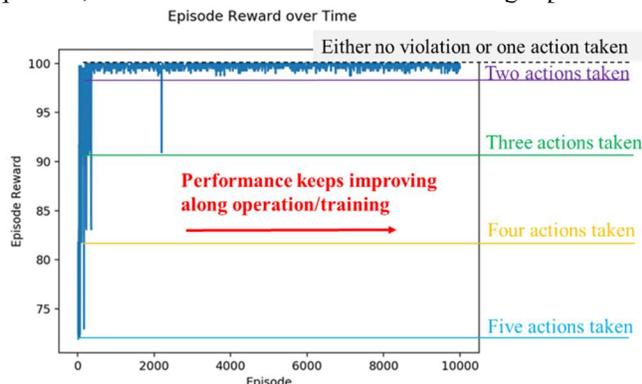

Fig.8. Performance of DRL Agent in the learning process using 10,000 episodes considering N-1 contingencies.

*C. Case III – Using Cconverged Agent with High Rewards*

In Case III, the definition of final reward for any episode is revised so that a higher reward, in the value of 200, is issued when the Agent can fix the voltage profile using only one action; if there is any voltage violation in the states, no reward is given. Using the updated reward definition and the procedures in Case II to train an Agent considering N-1 contingencies. Once the Agent is trained, it is tested on a new set of 10,000 episodes randomly generated with contingencies, by reducing exploration rate to a very small value. The test performance is shown in Fig. 9, demonstrating outstanding performance in autonomous voltage control for the IEEE-14 bus system. The sudden drop in reward around Ep 4100 is caused by exploration of a random action, leading to a few iterations before voltage problems are fixed.

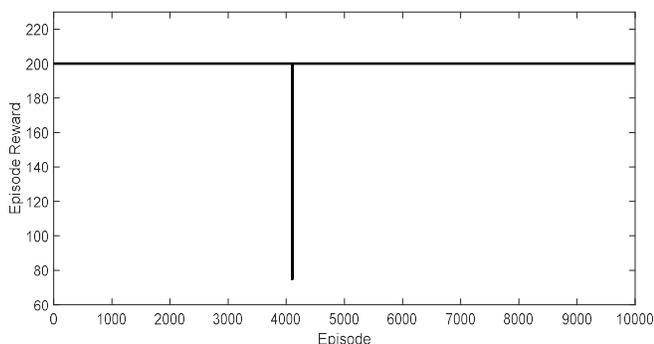

Fig. 9 DRL Agent performance on 10,000 episodes considering N-1 contingencies (exploration rate: 0.001, decay: 0.9, learning rate: 0.001) .

## V. CONCLUSIONS AND FUTURE WORK

To effectively mitigate voltage issues under growing uncertainties, this paper presents a novel paradigm, the **Grid Mind** framework, to use deep reinforcement learning for autonomous voltage control in grid operation. The architecture design, computational flow and implementation details are provided. The training procedures of DRL Agents are discussed in detail. The properly trained Agents can achieve the goal of autonomous voltage control with satisfactory performance. It is important to carefully tune the parameters of the Agent and properly set the tradeoff between learning and real-world application.

In future work, more control actions such as switchable shunt elements will be added to the DRL Agent learning process to provide more options to control voltage profiles. Larger power system models will be used to test the learning process and performance of DRL Agent.

ACKNOWLEDGMENT

The authors gratefully acknowledge the contributions of Xiao Lu and Haifeng Li from the State Grid Jiangsu Electric Power Company in China for their novel ideas and contributions.